\documentclass[fleqn,10pt]{wlscirep}
\usepackage{pdfpages}
\title{Evolution of Pairing Orders between Pseudogap and Superconducting Phases of Cuprate Superconductors}

\author[1,2,3]{Wei-Lin Tu}
\author[3*]{Ting-Kuo Lee}
\affil[1]{Department of Physics, National Taiwan University, Daan Taipei 10617, Taiwan}
\affil[2]{Laboratoire de Physique Théorique, IRSAMC, Université de Toulouse, CNRS, UPS, France}
\affil[3]{Institute of Physics, Academia Sinica, Nankang Taipei 11529, Taiwan}

\affil[*]{tklee@phys.sinica.edu.tw}

\begin{abstract}
One of the most puzzling problems of high temperature cuprate superconductor is the pseudogap phase(PG) at temperatures above the superconducting transition temperature in the underdoped regime. The PG phase is found by the angle-resolved photoemission spectra(ARPES) to have a gap at some regions in momentum space and a fraction of Fermi surface remained, known as Fermi arcs. The arc turns into a d-wave SC gap with a node below the SC transition temperature. Here, by studying a strongly correlated model at low temperatures, we obtained a phase characterized by two kinds of pairing order parameters with the total momentum of  the Cooper pair to be zero and finite. The finite momentum pairing is accompanied  with a spatial modulation of pairing order, i.e. a pair density wave (PDW). These PDW phases are intertwined with modulations of charge density and intra-unit cell form factors. The coexistence of the two different pairing orders provides the unique two-gaps like spectra observed by ARPES for superconducting cuprates. As temperature raises, the zero momentum pairing order vanishes while the finite momentum pairing orders are kept, thus Fermi arcs are realized. The calculated quasiparticle spectra has the similar doping and temperature dependence as reported by ARPES and scanning tunneling spectroscopy(STS). The consequence of breaking symmetry between x and y due to the unidirectional PDW and the possibility to probe such a PDW state in the PG phase is discussed. 
\end{abstract}

\DeclareUnicodeCharacter{2212}{-}
\begin{document}

\flushbottom
\maketitle
%
%
\thispagestyle{empty}

\section*{Introduction}

A long standing unresolved puzzle of the cuprate high temperature superconductors is the nature of pseudogap(PG) phase \cite{Vojta, Keimer}. Below the PG temperature $T^*$ there are experimental evidences  of breaking some crystalline symmetry \cite{Shekhter, Zhao}. Breaking of time-reversal symmetry with observation  of intra-cell magnetic moments has also been reported \cite{Bourges}. Many more new evidences suggest that this phase should be a nematic phase that breaks the four-fold rotation symmetry of the copper oxygen lattice \cite{Sato, Comin4, Wu2}. In particular there are many reports of the charge density waves(CDW) or spin density waves(SDW) in the SC and PG phases \cite{Yamada, Abbamonte, Comin, Comin2, Yazdani}.  Some of these are likely unidirectional hence without four-fold rotation symmetry. There are experimental evidences indicating the presence of fluctuating or short-range-ordered CDW in the PG phase \cite{Comin2, Comin3, Torchinsky}. Furthermore these CDW  orders were also observed in the superconducting  (SC) phase \cite{Comin3, Ghiringhelli, Neto, Hashimoto2, Blanco-Canosa, Neto2}. Once the CDW sets in and breaks four-fold symmetry \cite{Kohsaka}, the symmetry of pairing order in the SC phase of tetragonal crystal such as $Bi_2Sr_2CaCu_2O_{8+x}$ should not be expected as a pure $d$-wave as seen in experiments \cite{Kirtley, Tsuei}. Thus the formation mechanism of these density waves and its relations with SC and pseudopgap phases are of great interests.

Before the discovery of these density wave orders in the cuprates, the PG phase has already posed a number of unexplained puzzles. Below a characteristic temperature $T^*$ but higher than the SC transition temperature $T_c$, the excitation spectra showing a gap was first noticed by the relaxation rate of nuclear magnetic resonance \cite{Warren} and then by many other transport and spectroscopic measurements \cite{Timusk}. But the most direct observation of this gap structure was shown by the ARPES \cite{Marshall, Loeser, Ding}. The energy-momentum structure shows an energy gap appears near the boundary, or the antinodal region, of the two-dimensional Brillouin zone(BZ) of the cuprate. However there are four disconnected segments of Fermi surface near the nodal region, or $|k_x|=|k_y|=\pi/2$.  These segments called Fermi arcs have been reported to have their length shrink to zero \cite{Kanigel, Nakayama} when extrapolated to zero temperature. There are also results indicating that the arc length is not sensitive to temperature \cite{Loeser, Comin3}. Then it could also be part of a small pocket \cite{Doiron-Leyraud, Bangura}. This presence of finite fraction of Fermi surface is consistent with the Knight shift measurement \cite{Kawasaki} showing a finite density of states(DOS) after the superconductivity is suppressed. The full Fermi surface is recovered  either for temperature higher than $T^*$ or when doping increases beyond approximately 19$\%$ as the PG phase disappears. Below $T_c$ the gap at antinode merges with the SC gap. Also the ARPES spectra at the antinodal region does not have the usual particle-hole symmetry associated with traditional superconductors. This asymmetric antinodal gap onsets at $T^*$ and it persists all the way to the SC phase \cite{Hashimoto, He}. 

The phenomena of two gaps, one PG formed above the SC temperature $T_c$ and additional SC gap below, and all the exotic behavior associated with it has attracted many attentions as discussed in recent reviews \cite{Hashimoto, Huefner}. There are many theoretical proposals devoted to understand the PG as discussed in these review articles \cite{Vojta, Fradkin, Lee}. But so far it has been difficult to understand the temperature and doping dependence of the Fermi arcs, two gaps and other spectroscopic data, as well as its explicit relationship with the CDW orders and whether any of these are related with the Mott physics or the strong correlation.

However, there are growing evidences that these CDW are not a usual kind but are related to or could be a subsidiary order of the pair density wave(PDW). PDW is a state with spatial modulation of the pairing amplitude and it was first introduced by Larkin and Ovchinnikov \cite{Larkin} and by Fulde and Ferrell \cite{Fulde}. The pairing order $<C_{\textbf{k}+Q/2\uparrow} C_{-\textbf{k}+Q/2\downarrow}> =\Delta _Q(\textbf{k})$ has a center of mass momentum $Q$ and relative momentum $\textbf{k}$. Usual BCS pairing is for $Q=0$, or $\Delta_0(\textbf{k})$. There were quite a number of works proposing that PDW state might be responsible for the many observed exotic phenomena \cite{Podolsky, Chen, Himeda, Berg2, Berg3, Berg4} in both SC and PG phases. Many of the works used phenomenological models and weak coupling approaches \cite{Sachdev, Efetov, YWang, YWang2}, but some of the numerical works on microscopic models such as the Hubbard model and its low-energy effective $t-J$ model, have found strong evidences for such a state or states.  Actually there are many different kinds of PDW states that could be either unidirectional \cite{Yang3} or bidirectional like a checkerboard, which are tabulated in Ref. \cite{Tu} by us. For the unidirectional PDW state intertwined with CDW and SDW, so called the stripe state, was first proposed by the variational calculation for the $t-J$ model \cite{Himeda}. It could have the same sign of $d$-wave pairing on each site or pairing is in-phase \cite{Tu} so that the period of modulation of pairing is same as charge density but only half of the SDW. Or it could be the anti-phase stripe having two domains with opposite pairing sign so that the period of pairing modulation is twice of the charge density. The in-phase stripe was later shown \cite{Chou3} to be a stable ground state with half a hole in each period of CDW when a small electron-phonon interaction is included in the $t-J$ model. This half-doped stripe may be what was observed in neutron scattering \cite{Yamada} for the LBCO ($La_{2-x}Ba_xCuO_4$) family. 

The most strong theoretical support for the stripe state is from the recent numerical calculation by Corboz \textsl{et al.} \cite{Corboz}. By using the infinite projected entangled-pair states(iPEPS) method, they found that the $t-J$ model has several stripe states, with nearly degenerate energy as the uniform state. Even though the method has demonstrated its ability to obtain the lowest energies than all previous numerical methods, they still cannot pinpoint which state is the ground state. The results are quite consistent with the most recent numerical studies on the Hubbard model \cite{ZhengLin}. They found the stripe states have lower energies than the uniform SC state at 1/8 hole density and for $U/t=8$ and 12. The period of the PDW moves toward 4 or 5 lattice spacing as $U$ increases and this is more in line with result of the $t-J$ model. 

Besides the stripe state there is also the possibility of a PDW state coupled with CDW only and without SDW involved. Capello \textsl{et al.} \cite{Capello} have proposed such a state with  an uniform pairing order but it is not a pure $d$-wave order. Instead of proposing a possible state by conjecturing, we have solved a set of self-consistent equations derived from the renormalized mean-field theory(RMFT) \cite{Tu, Yang3} based on the generalized Gutzwiller approximation(GWA) \cite{Gutzwiller}. Of the many low energy solutions we found, there is a PDW state coupled with a uniform $d$-wave pairing order that explains a number of properties measured by the scanning tunneling spectroscopy(STS) on BSCCO($Bi_2Sr_2CaCu_2O_{8+x}$) and NaCCOC($Ca_{2-x}Na_xCuO_2Cl_2$) \cite{Hamidian}. This particular state, referred to as nodal pair density wave(nPDW) \cite{Tu, Choubey}, has pairing orders $\Delta_Q(\textbf{k})$, $\Delta_{-Q}(\textbf{k})$, for several $Q$ and also an uniform $\Delta_0(\textbf{k})$ that produces a nodal like local density of states(LDOS). The period of the CDW is about half of the PDW.  Furthermore, by including the Wannier function in our calculation to take into account the effect of oxygens that were neglected in the simple $t-J$ model \cite{Choubey}, we are able to compute the continuum local density of states of the nPDW. The energy dependence of intra-unit cell form factors and spatial phase variations of these states agrees remarkably well with the STS experiments \cite{Hamidian, Fujita2}. The success of this nPDW state to quantitatively explain the real space spectra measured by STS in the SC phase naturally leads us to study the spectra in momentum space measured by ARPES.

Instead of concentrating on the microscopic models, the Landau-Ginzburg free energy formalism is used to study the intricate relationship between PDW, CDW, and the uniform pairing order \cite{Berg, Berg2, Berg3, Berg4, Agterberg}.  By including phases of PDW, they could discuss vortex and dislocations as well as the phase diagram. They pointed out that PDW could be responsible for the PG phase. Some of the properties we shall discuss below are consistent with their results; however, they did not consider bond order as an independent field whereas we have shown \cite{Tu} that bond order with dominant intra-unit-cell form factor with $s'$ or $d$ symmetry are associated with different PDW states such as stripes or nPDW, respectively. Neither are most of the phenomenological approaches \cite{YWang2, Baruch}. Another interesting work by Lee \cite{Lee2} proposed the Amperean pairing originated from the gauge theory of the resonating valence bond(RVB) picture as the main mechanism for the formation of PDW and it is the dominant order in cuprates. This theory prefers to have bidirectional PDW to have similar gaps at antinodes $(\pi,0)$ and $(0,\pi)$. They also did not address the issue of bond orders.

In this paper the spectra associated with the nPDW state will be calculated first at $T=0$ as our previous work but with emphasis on the energy-momentum dependence of the quasiparticles. Then we will extend our calculations to finite temperatures. The GWA used in the RMFT is considered to be a good approximation at zero temperature. The energy scale imposed by the strong Coulomb repulsion, or Hubbard $U$, is much larger than the scale of room temperature. In addition, both the two main ``low'' energy scales, $t$ and $J$ about 3000$\sim$4000$K$ and 1200$K$, respectively, are also much larger. Hence we shall make an assumption that the GWA is reasonably accurate at low but finite temperatures. 

After the RMFT is transformed to solve for the self-consistent equations at finite temperatures, we found the average or net uniform pairing order parameter(UPOP) of the nPDW state decreases to almost zero at a ``critical" temperature $T_{p1}$. This new state still has incommensurate modulations of charge density, pair density and bond orders intertwined, and we shall denote it as incommensurate pair-density-wave(IPDW) state. Just as nPDW state this IPDW state also has the dominant intra-unit-cell $d$-form factors \cite{Tu, Choubey} and particle-hole asymmetry for the ARPES spectra \cite{He} at the antinodal region. The major difference with nPDW is the appearance of Fermi arcs and a substantial increase of DOS at Fermi energy but without UPOP. As temperature further increases to $T_{p2}$, there is no longer a solution of this state. The value of $T_{p2}$ increases sharply as doping is reduced. The DOS at Fermi energy increases only slightly between $T_{p1}$ and $T_{p2}$. The DOS also increases slightly with increasing doping. Comparing these results with experimental data on ARPES \cite{Hashimoto, He} and DOS deduced from Knight shifts \cite{Kawasaki}, we conclude that it is quite reasonable to take $T_{p1}$ as the SC transition temperature $T_c$ and $T_{p2}$ as a mean-field version of the PG temperature $T^*$ of the copper oxides. These issues will be discussed after the results are presented.

\section*{Results and Discussions}

All our results reported here are for the two-dimensional  $t-t'-J$ model shown in Eq.(\ref{Hamiltonian}), on a square lattice:
\begin{equation}
\begin{aligned}
H=-\sum_{i,j,\sigma}P_{G}t_{ij}(c^\dagger_{i\sigma}c_{j\sigma}+H.C.)P_{G}+\sum_{\langle i,j\rangle}JS_i \cdot S_j
\end{aligned}
\label{Hamiltonian}
\end{equation}
where nearest neighbor hopping $t_{ij}=t$ where $ij$ is set to be the nearest neighbor sites and $J$ is set to $0.3t$. $P_G=\prod_i(1-n_{i\uparrow}n_{i\downarrow})$ is the Gutzwiller projection operator, while $n_{i\sigma}=c^\dagger_{i\sigma}c_{i\sigma}$ stands for the number operator for site $i$.  Spin $\sigma$ is equal to $\pm$. $S_i$ is the spin one-half operator at site $i$. If $i$ and $j$ are nearest neighbor(next nearest neighbor), $t_{ij}=t$($t_{ij}=t'=-0.3t$). The effect of the projection operator $P_G$ is treated with a mean-field GWA; we replace the the constraint of forbidding the double occupancy of two electrons on the same site with Gutzwiller factors. Thus one can follow RMFT \cite{Yang3} to
find the various low energy states at $T=0$ as we have done in our previous works \cite{Tu, Choubey} focused on studying the STS. We shall first use these results to calculate the quasi-particle Green's functions and its spectra density. Then we will generalize the calculation to finite temperatures by assuming the Gutzwiller factors remain unchanged  as temperature is slightly raised. Discussion about the calculation is presented in the Methods section. 

Usually in our  theory there are four intertwined mean-field order parameters:  local AF moment $m_i^v$,  pair field $\Delta_{ij\sigma}^v$, bond order $\chi_{ij\sigma}^v$, and hole density $\delta_i$, where $i$ is a site position and $ij$ is the nearest neighbor bond, but in this paper we will not consider $m_i^v$. By including the Gutzwiller factors in the $t-t'-J$ model, the mean field Hamiltonian can be written as the BdG equations. After it is diagonalized, iterative method is used to achieve self-consistent solutions. If we are only interested in unidirectional density waves with modulation in x direction, we could exploit the translational invariance in y direction by switching the site index $i=(i_x, i_y)$ to $(i_x, k_y)$ by taking the Fourier transform in $i_y$ as shown in Ref. \cite{Choubey}, and this makes the numerical calculations much more efficient. As shown in Refs. \cite{Tu, Choubey}, there are many nearly energy degenerate solutions with commensurate or quasi-incommensurate modulations with several periods mixed together. The $d$-wave pair field at each bond could have the same sign or they have domains with opposite signs. Most of the inhomogeneous solutions have PDW intertwined with CDW and bond order. But if doped hole concentration is low in the very underdoped region, there are also solutions with spin density wave (SDW) intertwined with the other three orders. Readers can find detail discussions about these states and also bidirectional checker board solutions in Ref. \cite{Tu}. Here we will only consider nonmagnetic solutions.

\subsection*{Particle-hole asymmetry}

We will first discuss ARPES spectra for nPDW states \cite{Choubey} that have been  shown to provide quantitative agreement with STS experiments \cite{Hamidian} on BSCCO and NaCCOC. These states with incommensurate PDW, CDW, and bond order wave coexisting have a UPOP exhibiting a $d$-wave nodal like LDOS at low energy. Their energy dependence of the intra-unit-cell form  factors with $s$, $s'$ and $d$ symmetry  and the spatial phase difference agree well with the STS experiments.   The characteristics of a particular nPDW state obtained at hole concentration 0.125 for a $32\times32$ lattice is quite similar with our previous works in \cite{Tu, Choubey}. Figure S1 in the supplementary material(SM) shows the variation of  hole density and pairing order parameter as a function of positions(Fig. S1(a)) and their Fourier transform(Fig. S1(b)), the LDOS at several selected sites(Fig. S1(c)), and the the intra-unit-cell form factors for three symmetries(Fig. S1(d)). As shown in Fig. S1(b), The state has various periods mixed but mostly a period close to $4a$ in the x-direction for charge density modulation and $8a$ for pairing order modulation. 

\begin{figure}[ht]
\centering
\includegraphics[width=10.0cm]{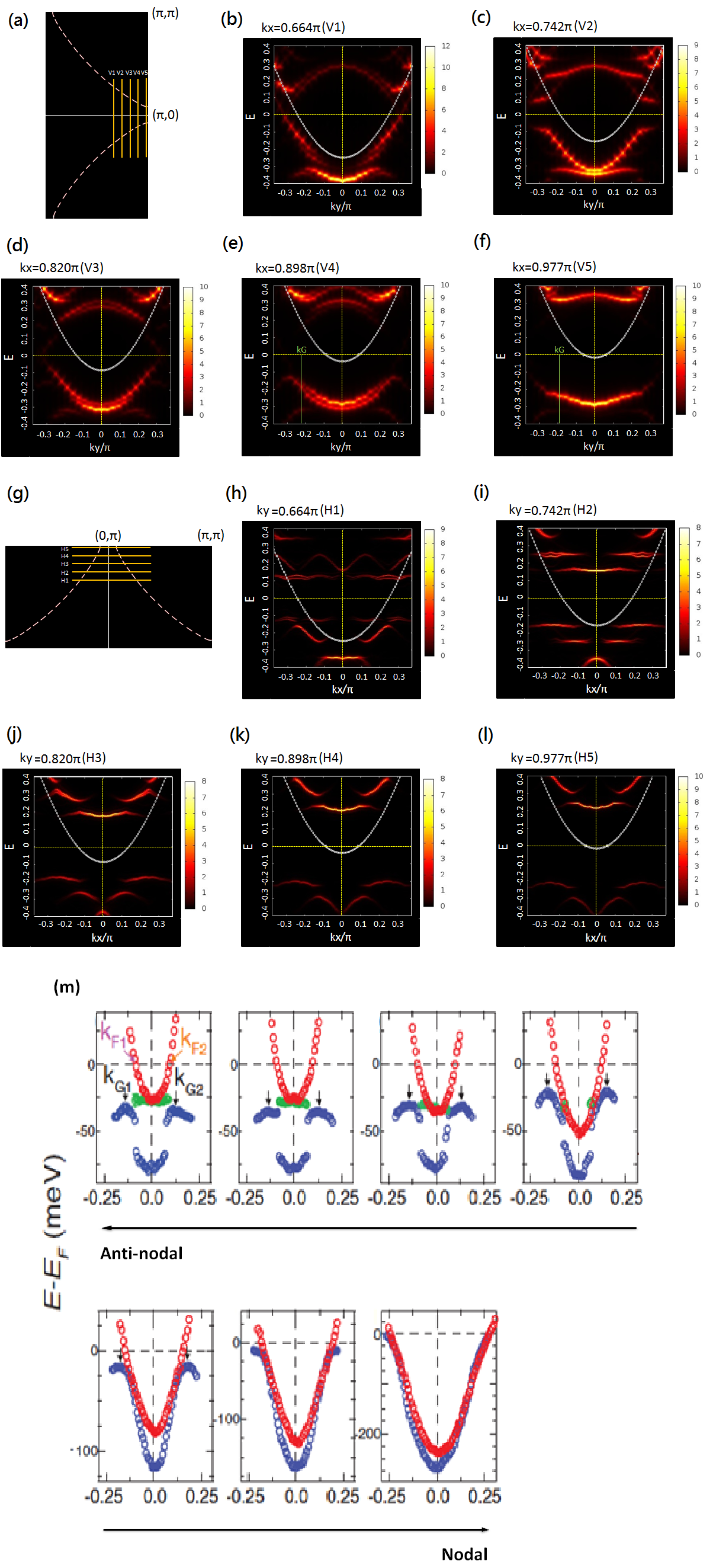}
\caption{}
\label{Fig.1}
\end{figure}

The spectral density $A (k_x,k_y,\omega)$ of the above state is calculated by using Eq.(\ref{spectra}) at $T=0$. We choose the width $\Gamma=0.01t$ unless specifically mentioned otherwise. In Fig. \ref{Fig.1}(a) we scan the momentum space near the antinodal region  $(k_x,k_y)=(\pi,0)$  by having 5 vertical cuts($V1$-$V5$)  perpendicular to the $k_x$ axis. The energy dependence of the spectral weight as a function of the y component of the wave vector($k_y$) for the five cuts are shown in Figs. \ref{Fig.1}(b)-\ref{Fig.1}(f). A very striking result is the complete lacking of particle-hole symmetry in the spectra as a BCS theory would have predicted. The white curves denote the dispersion of  a uniform  Fermi-liquid state(FLS) without pairing at dopant density 0.125. The curve crossing zero energy are Fermi momenta $k_F$. The five cuts show that near the nodal region $V1$ the gap at  $k_F$ is small and increases substantially approaching toward the antinodes $V5$. Most interesting part is the dispersion along each cut bends back after passing the   minimum energy gap, which is determined by looking at bands above and below Fermi energy. These back-bending momentum $k_G$ moves away from $k_F$ as momenta approaching antinodes. The momentum $k_G$ is determined by using the energy distribution curves(EDCs). Details are discussed in SM.  For the ARPES experiment, as demonstrated in Fig. \ref{Fig.1}(m), only the occupied states or the states with negative energies are measured, hence it cannot show the momenta with minimum gap but it can determine the back-bending momenta $k_G$. Indeed our result is very consistent with the experiments \cite{He} showing this particle-hole asymmetry which is very different from usual BCS superconductors that $k_G=k_F$. It was first pointed out by Lee \cite{Lee2} that the  difference between $k_F$ and $k_G$ and the way two approaches each other near nodal region is inconsistent with pure CDW either.

In the experiment, the spectra along $k_x$ direction is same as in $k_y$. This is likely due to the sample packed with x- and y- oriented short-range ordered unidirectional domains as seen in STS \cite{Hamidian}. But here we only have one unidirectional nPDW, hence spectra along $k_x$ and $k_y$ are different. In Fig. \ref{Fig.1}(g), we scan five horizontal cuts($H1$-$H5$) from near nodal region (h) to antinodal region (l). Comparing \ref{Fig.1}(b) with \ref{Fig.1}(h), we can see that there is no state at low energy for \ref{Fig.1}(h). In general the minimum gaps are the same for $H5$ cuts near $(0,\pi)$ and $V5$ cuts near $(\pi,0)$. It seems that the gap does not change much from \ref{Fig.1}(j) to \ref{Fig.1}(l) while it increases significantly from \ref{Fig.1}(d) to \ref{Fig.1}(f). The occupied bands are quite flat along the $k_x$ direction near $(0,\pi)$, and hence it is difficult to determine the bending vectors $k_G$. This kind of spectra near $(0,\pi)$ for a $x$-directional PDW is quite different from what one would expect for a pure CDW \cite{Lee2}. We also note that  energy gap value of $0.21t$ at the parallel direction (Fig. \ref{Fig.1}(I)) with respect to the modulation direction of density waves is about the same as in the perpendicular direction (Fig. \ref{Fig.1}(f)). This will be discussed further in the discussion section.

We have also considered anisotropy \cite{Yang3} in hopping $t_{x(y)}$ and $J_{x(y)}$ and if we decrease $t_x$ with respect to $t_y$ ($J_x/J_y = t^2_x/t^2_y$), then the nPDW state modulated in x-direction will no longer have a pure $d$-wave but a $s'+d$ wave. The energy gap determined by  the $H5$ cut near $(0,\pi)$ increases but the value of $V5$ cut near $(\pi,0)$ is reduced. We will discuss this further in the Discussion section later.    

\subsection*{Two Gaps in the SC phase}

The quasi-particle spectra in Fig. \ref{Fig.1} also show an energy gap increasing as momentum approaches antinodal region. To compare with  the result of  ARPES experiment we shall use the EDCs to determine the gap.  By taking a scan along a linear cut near the Fermi surface (the white curves in Fig. 1(a)), we can determine the k value that has the smallest energy difference from chemical potential. This will determine the energy gap at this $k$ value. Going from the nodal direction at $k_x=k_y$ to the antinodal region $(k_x,k_y)=(\pi,0)$, the energy gap is plotted as a function of  $|cos(k_x)-cos(k_y)|/2$ in Fig. \ref{Fig.3}(a) and (b) for nPDW states. In Fig. \ref{Fig.3}(a), at hole concentration $1/8$, the red curve is obtained for a  $16\times16$ lattice and green for $30\times30$. The vertical error bars are either determined by the width of the peak or by average of two nearby peaks with nearly the same magnitude. The horizontal error bars are due to the effect of discrete k values. The results are essentially the same for these two different lattice sizes and in fact, there is also not much difference with the solution of $60\times60$ lattice presented in \cite{Choubey}.  The vertical error bars are only slightly larger for the smaller $16\times16$ lattice. The slope of both curves increases as the momentum gets closer to the antinode. The dotted line in Fig. \ref{Fig.3}(a) indicates the linear fitting of a pure $d$-wave pairing gap $(\Delta_c|cos(k_x)-cos(k_y)|/2)$ near the nodal region. The fitted gap value $\Delta_c$ is much smaller than the gap at antinodal direction. The dashed line indicates a second gap near the antinodal region. Thus we have two different $d$-wave gaps near the node and antinode. 

Furthermore we can examine the variations of these gaps with dopant concentration. In Fig. \ref{Fig.3}(b) the gaps are plotted for nPDW states calculated for a $30\times30$ lattice for three hole concentrations: red for $\delta=0.1$, green for 0.125 and purple for 0.15. As hole concentration decreases, the gap at the antinodal region gets larger while deviation from the dotted line starts closer to the nodal $k$. The value of the gap at antinode could reach $0.2t$ or about 80 $meV$ just as in the experiments \cite{Hashimoto}. At larger doping these two $d$-wave gaps seem to approach each other as  a single gap which is expected in the usual BCS state. Due to the finite size effect, we cannot determine if the $d$-wave like gap for the three hole concentrations in Fig. \ref{Fig.3}(b) are exactly the same, but it looks close enough and with a value about $0.08t \sim$ 32 $meV$ for $t=0.4$ $eV$. In Fig. \ref{Fig.3}(a) and (b) we used a constant $\Gamma=0.01t$, but the result is insensitive to the choice of the $\Gamma$ in calculating the spectra density. This is very consistent with ARPES results shown in Ref. \cite{Hashimoto, Vishik} and we put alongside the figure of gap behavior of superconducting state from Ref. \cite{Hashimoto} in Fig. \ref{Fig.3}(c) for comparison. They found the gap value about $39$ $meV$ near the nodal region for several different hole concentrations.

So far we have discussed the gaps and spectral density of the superconducting nPDW states that were discussed previously in Ref. \cite{Tu, Choubey}. The state is quasi-periodic in the sense that it has several periods mixed but the dominant one is near $4a$. However it was recently proposed \cite{Mesaros} that the charge density modulation could be strictly $4a$ but the phase could jump with multiple values of $2\pi/4$ from its original perfect periodic values. This is the CDW with phase ``discommensuration"(DC) as defined by McMillan \cite{McMillan}. In the SM, Fig. S2 shows a nPDW state with $4a$ charge density modulation and phase DC for a $36\times36$ lattice at doping 0.125. The energy per site has  the same value as the previous nPDW state. Almost all the properties are very similar with the nPDW states discussed before \cite{Tu, Choubey}. The intra-unit-cell form factors are dominated by $d$-symmetry. The pair density modulation is mostly dominated by a vector $Q_p$, and the CDW has mainly a peak at $2Q_p$ and also a peak at $Q_p$. We will not discuss it here and detail can be found in SM.

\subsection*{Finite temperature IPDW states}

All the states discussed above are SC with a net $d$-wave UPOP. These were obtained by solving the mean-field BdG equations self-consistently at $T=0$. This also implies that we have a finite $\Delta_0(\textbf{k})$ as UPOP is same as average of the product of  $(cos(k_x)-cos(k_y))$ and $\Delta_0(\textbf{k})$. We recall that the strong correlation effect of Mott physics is originated from the very large on-site Coulomb repulsion or Hubbard $U$. This effect is translated into a Gutzwiller projection operator to prohibit double occupancy of electrons at each lattice site in the $t-t'-J$ model(Eq. (\ref{Hamiltonian})). By following the GWA \cite{Gutzwiller}, we replace these projection operators  by  Gutzwiller factors, which are functions of hole density. Here we will make an intuitive assumption that these Gutzwiller factors remain unchanged at temperatures much smaller than  the relevant energy scale $t$ and $J$ which are of order 0.4 $eV$ and 0.12 $eV$, respectively. Thus the BdG equations are easily generalized to finite temperatures and we could again find self-consistent solutions at finite $T$. Details are discussed in the Methods section. Here we will present the results.

In Fig. \ref{Fig.4}, the average or net UPOP calculated for a  lattice of $30\times30$ with doping 0.125, 0.15, and 0.16 are plotted as a function of $T/t$, shown in green, blue and red marks. $T_{p1}$ and $T_{p2}$ for the three hole concentrations are also denoted. For simplicity $T_{p1}$ is determined when the magnitude of UPOP reaches about 0.001. The three curves are quite similar except that near $T=0$, $x=0.16$ has the largest pairing order and also the largest value of $T_{p1}$; however its $T_{p2}$ is the smallest. The meaning of $T_{p2}$ where no PDW exists becomes more clear when we examine its doping dependence. For $T > T_{p2}$, the phase is actually a uniform $d$-wave state without modulations of charge and pairing. In previous work \cite{Tu}, we already showed that the nPDW state and all other CDW or SDW states have a slightly  higher energy than the uniform BCS state for the $t-J$ and $t-t'-J$ model. If we include other interactions like long range Coulomb interaction or a weak electron-phonon interaction \cite{Chou2, Chou3}, these density wave states could become lower in energy. Even if we only consider $t-t'-J$ model, these nPDW states are stable solutions at local minimum, until they no longer exist at $T_{p2}$ as shown in Fig. \ref{Fig.4}. Between $T_{p1}$  and $T_{p2}$ this new PDW state has essentially a zero UPOP or $\Delta_0(\textbf{k})=0$ but still finite $\Delta_Q(\textbf{k})$. It is an incommensurate PDW(IPDW) state, or the FFLO \cite{Larkin, Fulde} state, intertwined with modulation of charge density and bond orders.

The pattern of  pairing order  at each bond and hole density  for an IPDW state with $\delta=0.15$ at $T=0.035t$ is shown in Fig. \ref{Fig.5}(a). The hole density is maximum at sites, e.g. 2, 6 and 10 at the pairing boundary where the paring order changes sign. The  LDOS at a few selected sites are shown in Fig. \ref{Fig.5}(b). There is  a finite constant LDOS near  zero energy and it is not nodal like as the usual $d$-wave SC and nPDW states \cite{Tu, Choubey}. The Fourier transform of the intra-unit-cell form factor, hole density $\delta_i$ and pair field $\Delta_{ij\sigma}$ are shown in Figs. \ref{Fig.5}(c)-\ref{Fig.5}(e),  respectively. the definition of form factors are \cite{Choubey}:
\begin{equation}
\begin{aligned}
&d(\textbf{q})=FT(\widetilde{\chi}_{i,i+\hat{x}}-\widetilde{\chi}_{i,i+\hat{y}})/2\\
&s'(\textbf{q})=FT(\widetilde{\chi}_{i,i+\hat{x}}+\widetilde{\chi}_{i,i+\hat{y}})/2\\
&s(\textbf{q})=FT(1-\delta_i)
\nonumber
\end{aligned}
\end{equation}
where $FT$ refers to the Fourier transform and $\widetilde{\chi} _{i,j}$ denotes the bond order (Eq.(\ref{parameters})) \cite{Tu} with the mean value subtracted to emphasize the modulating component. $\hat{x}$($\hat{y}$) means one lattice displacement in x-(y-) direction. Both the modulation wave vector of the bond order wave and CDW, shown in Figs. \ref{Fig.5}(c) and \ref{Fig.5}(d), respectively, are $2Q_p=0.52\pi/a$, while the pairing modulation is dominated by $Q_p=0.26\pi/a$ as shown in Fig. \ref{Fig.5}(e). Most of the properties of this IPDW state are similar with nPDW state except three distinctions: a negligible net UPOP, a finite Fermi arc as shown in Fig. \ref{Fig.5}(f) and $FT$ of charge density has no peaks at $Q_p$ \cite{Berg3, Berg4}. The color legend represents the spectral weight of these $k$-points on the arc in Fig. \ref{Fig.5}(f). Notice the arc is asymmetric with  respect to exchanging  $k_x$ and $k_y$  as the modulation along x-direction breaks the $x$ and $y$ symmetry. At the antinodal region the Fermi surface is gapped out similarly as the nPDW states.

In Fig. \ref{Fig.6}(a), $T_{p1}$ and $T_{p2}$ are plotted as a function of doped hole concentration with the blue triangles and diamonds respectively. We also plotted the PG phase temperature $T^*$ determined from the NMR measurement \cite{Kawasaki} for $Bi_2Sr_{2-x}La_xCuO_{6+\delta}$ in red color by taking $t$ to be 0.4 $eV$. $T_{p1}$ follows a dome shape and has a maximum at hole concentration 0.16. The steep suppression of $T_{p2}$ with doping is similar with the PG temperature $T^*$, and the values are also close if we reduce $T_{p2}$ by about a factor of 2. This is not surprising as we have neglected the quantum fluctuation effect in this mean field theory \cite{Yang4}, and we also have assumed the Gutzwiller factors to have no $T$ dependence. Also note that so far we have only considered long-range-ordered solutions and have neglected solutions of random x- and y- oriented domains with short-range IPDW states. Since IPDW state has a Fermi arc as shown in Fig. \ref{Fig.5}(f), we expect the gap should vanish at the Fermi surface near the nodal region. In Fig. \ref{Fig.6}(b) the gap value along the Fermi surface is plotted for $T=0$ (red squares) and $T=2T_{p1}$(green squares). As a comparison, the IPDW state for 0.11 is also shown with the purple marks. Error bars are discussed in SM. This is very close to what is measured on BSCCO by ARPES \cite{Vishik}. The gap at antinode is essentially unchanged when the state changes from nPDW to IPDW.  This is not surprising, since the antinode is still much larger than the temperature.

A very important property of the PG phase is the Knight shifts measured by NMR \cite{Kawasaki}; it shows that the  DOS in the PG phase  increases slowly with doping but is less than half of the DOS of the normal state for $T>T^*$ until it is near the critical doping about 0.2 where the PG phase disappears.  It is also found that the Knight shift or DOS varies with temperature by less than 10$\%$ during the PG phase.

In the inset of Fig. \ref{Fig.6}(c), DOS is plotted as a function of temperature for three hole concentrations.  Here we have assumed the width $\Gamma$, used in the spectra density calculation, is of the form $\Gamma = 0.25 \sqrt{E^2+T^2}$ \cite{Alldredge}.  The DOS is calculated at zero energy(within an energy range of $\pm 0.004t$) by averaging the LDOS at all sites. The DOS are all quite small and almost the same at $T = 0$ but it increases significantly at $T_{p1}$. The DOS  values between $T_{p1}$  and $T_{p2}$  increase with doping. This is likely due to the fact that the length of Fermi arc  increases with doping. The variation of DOS with $T$ between $T_{p1}$ and $T_{p2}$ for these three hole concentrations are also near 10$\%$ as in experiments. In Fig. \ref{Fig.6}(c), the ratio of DOS between the IPDW states and the FLS  is plotted as a function of dopant concentration at $T=0.035t$. Not only the doping dependence is very close to the experimental data \cite{Kawasaki} shown as red symbols, the values are also close to the measured results. It is difficult for us to obtain solutions above dopant concentration 0.17 as $T_{p2}$ and $T_{p1}$ are very close(Fig. \ref{Fig.6}(a)). When the dopant concentration is above 0.18, we have no nPDW solution at $T=0$ and no IPDW state at finite $T$. Thus we would recover the full Fermi surface and the relative DOS should be 1. For real materials this happens at concentration 0.2 instead of 0.18.

The quasi-particle spectra of the IPDW state is very similar with what is shown in Fig. \ref{Fig.1} for nPDW state. The back-bending momentum $k_G$ moves away from $k_F$ approaching the antinode.

\section*{Conclusion}

Assuming that the Gutzwiller factors which take into account the renormalization effect of the strong correlation physics could have very small temperature dependence below room  temperatures, we then generalize the renormalized mean-field theory to finite temperatures to study the prediction of the $t-t'-J$ model. 

At low-temperature SC phase with a finite UPOP, a special self-consistent solution, the nPDW state first found by us in \cite{Tu}, is shown to have the two $d$-wave pairing gaps as found by the ARPES. The smaller the doping, the larger is the gap magnitude at antinodes, but  the nodal gaps are almost same for different dopings. The larger particle-hole asymmetry reported near the antinodal region is also well produced by the calculated spectra function. This nPDW state has a very special property that although it has a one-dimensional structure, the net pairing order or UPOP still has the four-fold $d$-wave symmetry. It is quite amazing that although the pairing value at each bond looks quite random (Fig. S1(a) in SM), its average has an exact $d$-wave symmetry. The spectra at antinodes $(\pi,0)$ and $(0,\pi)$ are quite different but the values are close to each other \cite{Gap}. Combining together with previous works \cite{Tu, Choubey} comparing our calculated LDOS and local spectra with the STS measurement, we  have obtained a very consistent picture about experimental data for both spectra in real space and in momentum space for the superconducting phase.

Here it is worthwhile to make a special discussion that the nPDW state we chose is among many possible solutions with different periods. Fortunately most of them examined by us have very similar properties except that the periods of modulations could be different.   In terms of energy the uniform $d$-wave SC state is the ``true" ground state of the $t-t'-J$ model within our RMFT. However, as we mentioned earlier and before \cite{Tu}, whenever other weak interactions, such as electron-phonon interaction, nearest neighbor or long range Coulomb force and impurities or defects, are added to the model \cite{Chou2, Chou3}, the nPDW state could be stabilized. Even if we only consider pure $t-t'-J$ model, these states are in their local minimum. Thus we could study its low energy excitations. At the end of this section, we will discuss the case with different hopping rates $t_x$ and $t_y$ along x and y directions. Then nPDW is stabilized as the ground state if the difference between $t_x$ and $t_y$ is large enough. 

Another thing we like to point out is that experiments just like our theory also  have found different kinds of CDW states. For the $La_{2-x} Ba_x CuO_4$ family, the period of CDW decreases with doping while it increases for YBCO and BSCCO \cite{Comin}. These two are called CDW1 and CDW2, respectively,  in the review article \cite{Fradkin}. There is also CDW3 or the magnetic field induced CDW. In this work we only  concentrated on  CDW2, which has no magnetic component. We believe CDW1 is probably the stripe state \cite{Chou3}. 

When the temperature is raised, the net UPOP in the nPDW state begins to decrease and it becomes negligible at certain temperature $T_{p1}$. This behavior also supports our assumption that these states are at a local minimum. Then it changes into an IPDW state that still has incommensurate modulations of charge density, bond order and pairing order $\Delta_Q(\textbf{k})$ for finite Q and with $\Delta_0(\textbf{k})=0$. Magnitude and modulation periods of all these three orders are quite similar to the nPDW state except that the FFT of the charge density does not have a peak at the wave vector of the  pairing modulation as seen in nPDW \cite{Berg3}.  In IPDW state, the modulation momentum of charge is twice of pairing. These states vary gradually with temperature until it reaches a higher temperature $T_{p2}$ and there is no longer a solution with modulations of pairing. Quite unexpectedly $T_{p2}$ actually decreases sharply as doping increases.  Fig. \ref{Fig.6}(a) shows that $T_{p2}$ is proportional to the PG temperature $T^*$ with an overestimation of at most a factor two for its values. This is quite satisfactory for a simple mean-field approach like ours. More discussion about this is given below.

Furthermore our analysis shows that the IPDW state near nodal region has a Fermi arc with a fraction of DOS of the full Fermi surface when there is no pairing. There is still a large gap at the antinodal region as shown in Fig. \ref{Fig.6}(b). The DOS  or the length of Fermi arc increases with dopant concentration just as what were seen by ARPES \cite{Hashimoto} and NMR \cite{Kawasaki}. 

In our calculation we obtain the uniform $d$-wave SC state at $T$ greater than $T_{p2}$. However as we mentioned earlier, this could be a consequence that we actually are at a local minimum and uniform SC state is the global minimum solution.  We believe that if we consider solutions composed of randomly packed x- and y- oriented domains of these IPDW states, its large entropy would have a lower free energy than that of the uniform SC state. Thus the reappearance of the uniform $d$-wave SC state at high $T$ is indicating the limit of accuracy of our mean-field theory and it probably has no physical significance. Here we notice that much more accurate numerical work \cite{ZhengLin} than our mean-field result for Hubbard model  at dopant 0.125 shows that uniform state  is not the ground state. The stripes including  SDW in addition to PDW and CDW are possible ground states for $U/t=12$ or less.  For larger $U$ as is for $t-J$ model,  antiferromagnetism is weaker and the nature of ground state is yet to be settled.  

It should be emphasized that  the IPDW state with finite pairing order $\Delta_Q(\textbf{k})$ is also a SC FFLO \cite{Larkin, Fulde} state with finite momentum pairing if there is a  phase coherence. But actually there maybe solutions with disordered fluctuating domains \cite{Huang} with different charge density, phases and periods, etc. Variational Monte Carlo calculations have shown \cite{Chou4} that random stripe domains could be very competitive in energy in comparison with the long-range-ordered state. Furthermore the short-range-ordered domains of these IPDW states will have larger entropy and lower free energy. The PG phase is known to have strong vortex fluctuations \cite{Xu, Anderson2}. The inclusion of phases  for these PDW states and their coupling with vortices \cite{Berg3, Berg4, Agterberg} will provide a better and wholistic description of the PG phase. However, the PDW described here, we believe, should still be a basic entity included in these considerations to account for the spectra measured by experiments.

Another important issue we have not addressed is the effect of magnetic field on the PDW.  Magnetic field-induced unidirectional CDW states have been reported below and above $T_c$ for YBCO \cite{Chang, Wu3, Blanco-Canosa2}. Some are even long-range ordered in 3D \cite{Gerber}. Somewhat different results are found in BSCCO family.  Recent experiment on BSCCO has found bidirectional PDW or checkerboard of 8 unit cell period existing inside the vortex halo \cite{Edkins}. For $Bi_2Sr_{2-x}La_xCuO_6$, NMR measurement \cite{Kawasaki2} shows that an in-plane magnetic field of 10$T$ is enough to induce long-range ordered CDW without spin components in the PG phase. Since such a small in-plane field does not have much effect on our nPDW or IPDW states, we believe these states are the ones observed in $Bi_2Sr_{2-x}La_xCuO_6$. This is supported by the good agreement achieved in Fig. \ref{Fig.6}(c) between the calculated DOS of our IPDW states and the Knight shift measurement in the PG phase by suppressing SC phase with high field \cite{Kawasaki}.

One of the surprises we found is that without adjustable parameters in our calculations,  we are able to get many quantities very close to experimental values and also have very good agreement with very sophisticated numerical works that go much beyond mean-field theory. Considering we are doing a mean-field calculation, this is even more surprising. One main reason could be that the GWA is really effective in catching the strong-correlation physics of the $t-J$ model. As discussed in our previous paper \cite{Tu}, the doping dependence of the GW factors are very important in obtaining these solutions. If we set all GW factors to 1, we are not able to find any of these PDW solutions. Based on this premise, we can now provide a very simple picture about the cuprate phase diagrams. Starting at half-filling, the model has the RVB proposed by Anderson \cite{Anderson} dominate in the Mott insulator. RVB has the $d$-wave  pairing and bond order intertwined. But without charge present both of them are actually the variational parameters or hidden orders we defined in Eq. (\ref{parameters}). When holes are doped into the lattice, RVB tends to  localize the charges to prohibit its fluctuation. Once the localization has failed possibly after the antiferromagnetism is destroyed by the dopant, the system starts to form these unidirectional PDW states which has charge density intertwined with RVB (pairing and bond order). These states have a gap in the antinodal region and in the nodal region a Fermi arc with only a fractional DOS survived. When there is too much doping that these density waves can no longer be viable, then we lose the Mott physics and recovered a FLS \cite{Badoux}.  These states then develop an average uniform SC pairing order at lower temperatures although it is relatively small in comparison with large magnitude of pairing modulation. Of course, the phase fluctuation will become more important as temperature rises \cite{Fradkin, Berg4} and mean field results will be revised. 

The theory we propose depends on the presence of a PDW state in the PG phase. There is a way to test this hypothesis besides the possibility of using STS \cite{Hamidian}, which has to worry about the rapid pairing phase variation in a few lattice spacing and also the measurement being most likely at a higher temperature. For a PDW state in $x$-direction, the magnitude of the gap in the $y$-antinode $(0,\pi)$ is about the same \cite{Gap} as the gap in the $x$-antinode $(\pi,0)$ as shown in Fig. \ref{Fig.1}. This is contrary to what one would expect if we only have  a pure CDW in the $x$-direction.  Then the gap opening due to zone folding should be larger  in the folding direction. The $x$ and $y$ asymmetry of the  Fermi arc, shown in Fig. \ref{Fig.5}(f), may be used to distinguish the arc from part of the Fermi pocket \cite{Doiron-Leyraud, Bangura}. We can also examine the particle-hole  asymmetry in the PG phase. IPDW will have very similar result as the nPDW state measured by ARPES \cite{He}. Particle-hole asymmetry should be observed away from the Fermi arc. This could be a sign for the presence of finite momentum Cooper pairs \cite{Lee2}.

It has been shown that a magnetic field about $10$ $T$ is enough to induce a long-range ordered CDW or PDW \cite{Gerber,Kawasaki2}. Since $10$ $T$ is quite small, it may be possible to generate the long-range order by having a thin tetragonal single layer cuprate deposited on a strained substrate. We have looked at the case with the hopping rate in the $x$ direction $t_x$ less than $t_y$ in the $y$ direction. The preliminary result shows that the energy of nPDW state for doping concentration $x=0.08$ is now lower than the uniform $d$-wave SC state if $t_x$ $<$ 0.84 $t_y$. This is consistent with previous work on stripe states at $x=0.125$ \cite{Yang3}. But here we   probably overestimate the strength of the uniform state. In real material a small difference between $t_x$ and $t_y$ might be enough to stabilize an IPDW/nPDW.  Since the x-directional nPDW has a much lower energy than the y-directional nPDW, the system is likely to be dominated by only x-directional nPDW, and a unidirectional IPDW at $T>T_c$. It may also be possible to have a phase coherent IPDW state in a small temperature window that will be a truly new phase. Even without invoking $t_x < t_y$, as shown in Fig. \ref{Fig.1},  the spectra  near $(\pi,0)$ and $(0,\pi)$ are very different. Now with $t_x < t_y$, the UPOP has the $s'+d$ symmetry with pairing in x direction larger than in y. On the other hand, the energy gap near the x-direction antinode $(\pi,0)$ is getting smaller as strain increases, while the gap near $(0,\pi)$ becomes larger. Thus IPDW in the PG phase could be detected with ARPES in this system.

\section*{Methods}

Following the idea of Gutzwiller \cite{Gutzwiller} and works of Himeda and Ogata \cite{Himeda2, Ogata}, we replace the projection operator($P_G$) in the $t-t'-J$ Hamiltonian in Eq.(\ref{Hamiltonian}) with the Gutzwiller renormalization factors. The renormalized Hamiltonian now becomes
\begin{equation}
\begin{aligned}
H=&-\sum_{i,j,\sigma}g^t_{ij\sigma}t_{ij}(c^\dagger_{i\sigma}c_{j\sigma}+H.C.)\\
&+\sum_{\langle i,j\rangle}J\Bigg [ g^{s,z}_{ij}S^{s,z}_i S^{s,z}_j+g^{s,xy}_{ij}\Bigg(\frac{S^{+}_i S^{-}_j+S^{-}_i S^{+}_j}{2}\Bigg)\Bigg] \\
\end{aligned}
\label{Hamiltonian2}
\end{equation}
where $g^t_{ij\sigma}, g^{s,z}_{ij}$, and $g^{s,xy}_{ij}$ are the Gutzwiller factors, which are dependent on the values of hole density $\delta_i$, which is one of our variational parameters along with pair field $\Delta_{ij\sigma}^v$ and bond order $\chi_{ij\sigma}^v$: 
\begin{equation}
\begin{aligned}
&\Delta_{ij\sigma}^v=\sigma \langle\Psi_0 | c_{i\sigma}c_{j\bar{\sigma}} | \Psi_0 \rangle\\
&\chi_{ij\sigma}^v=\langle\Psi_0 | c^\dagger_{i\sigma}c_{j\sigma} |\Psi_0 \rangle\\
&\delta_i=1-\langle\Psi_0 | n_i |\Psi_0 \rangle
\end{aligned}
\label{parameters}
\end{equation}
where $| \Psi_0 \rangle$ is the unprojected wavefunction. The superscript $v$ is used to denote that these quantities are different from the real physical quantities for comparison with the experiments. The factors are given as
\begin{equation}
\begin{aligned}
&g^t_{ij\sigma}=g^t_{i\sigma}g^t_{j\sigma}\\
&g^t_{i\sigma}=\sqrt{\frac{2\delta_i}{1+\delta_i}} \\
&g^{s,xy(z)}_{ij}=g^{s,xy(z)}_i g^{s,xy(z)}_j\\
&g^{s,xy(z)}_i=\frac{2}{1+\delta_i}\\
\end{aligned}
\end{equation}
The more general forms of these factors with magnetic moment involved can be found in \cite{Tu, Yang3, Choubey, Christensen}. In this paper we only consider paramagnetic PDW. We then solve for the BdG equation and perform the iteration until the differences of consecutive variational order parameters are smaller than $10^{-3}$ and sometime  $10^{-4}$ to verify the $d$-wave order more accurately. For an nPDW state like in Fig. \ref{Fig.1}, we will have 80 variables to be solved self-consistently.

Because we are going to investigate the features in $k$ space, it is necessary to apply the supercell calculation \cite{Schmid}. For each cell we have $N_x  \times N_y$ sites and the total number of cell is $M_c=M_x \times M_y$. Our Hamiltonian is then reduced from $2M_xN_x \times 2M_yN_y$ to $M_x \times M_y$ matrix equation each with lattice size $2N_x \times 2N_y$. The self-consistent solutions now have to be carried out for each cell. The spectral weight can be written with our wave function $(u,v)$ as:

\begin{equation}
\begin{aligned}
A(k,\omega)=&\frac{1}{N}\sum_{ij,n+}f(-E_n)(e^{i\textbf{k}\cdot (\textbf{r}_i-\textbf{r}_j)}g^t_{ij\uparrow}u^{\textbf{K}\ast}_{i,n}u^{\textbf{K}}_{j,n}\delta(\omega-E_n)\\
&+e^{i\textbf{k}\cdot (\textbf{r}_j-\textbf{r}_i)}g^t_{ij\downarrow}v^{\textbf{K}}_{i,n}v^{\textbf{K}\ast}_{j,n}\delta(\omega+E_n))\\
&+\frac{1}{N}\sum_{ij,n-}f(E_n)(e^{i\textbf{k}\cdot (\textbf{r}_i-\textbf{r}_j)}g^t_{ij\uparrow}u^{\textbf{K}\ast}_{i,n}u^{\textbf{K}}_{j,n}\delta(\omega-E_n)\\
&+e^{i\textbf{k}\cdot (\textbf{r}_j-\textbf{r}_i)}g^t_{ij\downarrow}v^{\textbf{K}}_{i,n}v^{\textbf{K}\ast}_{j,n}\delta(\omega+E_n))
\end{aligned}
\label{spectra}
\end{equation}
where $\textbf{k}=\textbf{k}_0+\textbf{K}$ while $\textbf{k}_0=2\pi(\frac{n_x}{N_x},\frac{n_y}{N_y})$ where $n_x \in [-N_x/2+1,N_x/2], n_y \in [-N_y/2+1,N_y/2]$, and $\textbf{K}=2\pi(\frac{n^c_x}{M_xN_x},\frac{n^c_y}{M_yN_y})$ where $n^c_x \in [0,M_x-1], n^c_y \in [0,N_y-1]$. $f(E_n)$ is the Fermi-Dirac distribution and $n+$($n-$) means summation over positive(negative) energies. $\delta(\omega-E_n)$ is the Lorenzian and has the following form:
\begin{equation}
\begin{aligned}
\delta(\omega-E_n)=\frac{1}{\pi}\frac{\Gamma}{\Gamma^2+(\omega-E_n)^2}
\end{aligned}
\end{equation}

Since a lot of our patterns are unidirectional, we can exploit the translational invariance in y-direction assuming that the modulation is in x-direction to reduce the calculation time. By transforming our original creation/annihilation operators into those with basis of $(i_x,k)$:

\begin{equation}
\begin{aligned}
c^\dagger_{i,\sigma}=\frac{1}{\sqrt{N}}\sum_{k}c^\dagger_{i_x,\sigma}(k)e^{-ikR_{i_y}}
\end{aligned}
\end{equation}
we could translate our Hamiltonian as in a 1D lattice. With this transformation, we are able to perform the calculation for lattice size two times larger. For the symbols above, $N$ represents the lattice size in y-direction, $R_{i_y}$ is the y component of the original lattice vector $i$, and $c^\dagger_{i_x,\sigma}(k)$ is the creation operator in this quasi-1D system for momentum $k$. Details could be found in Ref. \cite{Choubey}.

The UPOP plays an important role in our work. To obtain it, we first calculate $\overline{\Delta_x}$ and $\overline{\Delta_y}$:
\begin{equation}
\begin{aligned}
&\overline{\Delta_x}=\sum_{\textbf{K}}\sum_i^{N_x} \Delta^{\textbf{K}}_{ii+\hat{x}} /N_x/M_c \\
&\overline{\Delta_y}=\sum_{\textbf{K}}\sum_i^{N_x} \Delta^{\textbf{K}}_{ii+\hat{y}} /N_x/M_c 
\end{aligned}
\end{equation}
After we obtain the averaged pairing values in x and y direction, we can then calculate UPOP:
\begin{equation}
\begin{aligned}
UPOP=\frac{|\overline{\Delta_x}|+|\overline{\Delta_y}|}{2}
\end{aligned}
\end{equation}

\bibliography{Tu-Lee-SciRe}

\begin{thebibliography}{10}
\bibitem{Gap}
 \bibinfo{note}{Gap values near the two antinodes my differ by 10$\%$ for
  different nPDW states. This is probably the accuracy of the mean-filed
  theory}.
\end{thebibliography}

\section*{Acknowledgements}

TKL likes to thank P. A. Lee, P. W. Anderson , F. C. Zhang, N. Nagaosa, and G. Margaritondo for valuable comments. The authors are grateful for the support of Taiwan Ministry of Science and Technology Grant 106-2112-M-001-020 and 105-2917-I-002 -010. Part of calculation was supported by the National Center for High Performance Computing in Taiwan.

\section*{Author contributions statement}

TK.L. conceived the original idea. W.T. and TK.L. provided the theoretical understanding and wrote the paper together.

\section*{Additional information}
\subsection*{Competing Interests}The authors declare no competing interests.

\section*{Figure Legends}

\subsection*{Figure 1}

The quasiparticle spectra of a nPDW state calculated in a $32\times32$ lattice for hole concentration 0.125: (a) the vertical cuts ($V1$-$V5$) denote the y component of the momentums scanned from (b)(near nodal region) to (f)(anti-nodal region). (b)-(f): quasiparticle spectra weight for each cut as a function of $k_y$ with a fixed $k_x$ value shown above each figure. (g) the horizontal cuts ($H1$-$H5$) denote the x component of the momentums scanned from (h)(near nodal region) to (l)(anti-nodal region). (h)-(l): quasiparticle spectra weight for each cut as a function of $k_x$ with a fixed $k_y$ value shown above each figure. (m) Spectra figures from the ARPES results \cite{He} for comparison. Notice that the sequence of showing these figures goes from the anti-nodal to nodal direction, which is different from that of \ref{Fig.1}(b) to \ref{Fig.1}(f).

\subsection*{Figure 2}

The gap value evolving from nodal to anti-nodal region for nPDW for (a) two different lattice sizes at doping 0.125; (b) different doping levels but same size(30$\times$30). Red, green and purple lines are just guides for the eyes. The black dotted(dashed) line is a fitted pure d-wave(antinodal) gap with gap size about $0.075\sim 0.08t$($0.2t$). The Gaussian width used here is $\Gamma=0.01t$. (c) Gap behavior from ARPES within the superconducting dome \cite{Hashimoto}.

\subsection*{Figure 3}

UPOP vs temperature for $\delta=0.125$, $0.15$, and $0.16$. $T_{p1}$ and $T_{p2}$ for each case are marked with different dotted lines of the same colors. The lattice size is $30\times30$.

\subsection*{Figure 4}

Properties of IPDW. (a) The real space modulation of IPDW. The red and black numbers on each bond denote the values of pairing order and the number at each site (black dots) is the hole density. (b) The LDOS for sites near the domain wall(2, 6, 9, 14 in 5(a)) and in the middle of nearby domain walls(1, 4, 8, 15 in 5(a)). (c) Different form factors and (d)(e) Fourier transform of hole density(d) and pairing order(e). The red vertical dashed lines mark $|q|=0.5\pi/a$ corresponding to period $4a$. Quasiparticle spectra with zero energy in k space for IPDW in $30\times30$ lattice sites at $T=0.035t$ are shown in (f) for $\delta$=0.15. The cyan dotted curve is the Fermi surface of Fermi liquid state with the same doping level. $\Gamma$ used here is equal to $0.25 \sqrt{E^2+T^2}$ \cite{Alldredge}.

\subsection*{Figure 5}

\begin{figure}[ht]
\centering
\includegraphics[width=\linewidth]{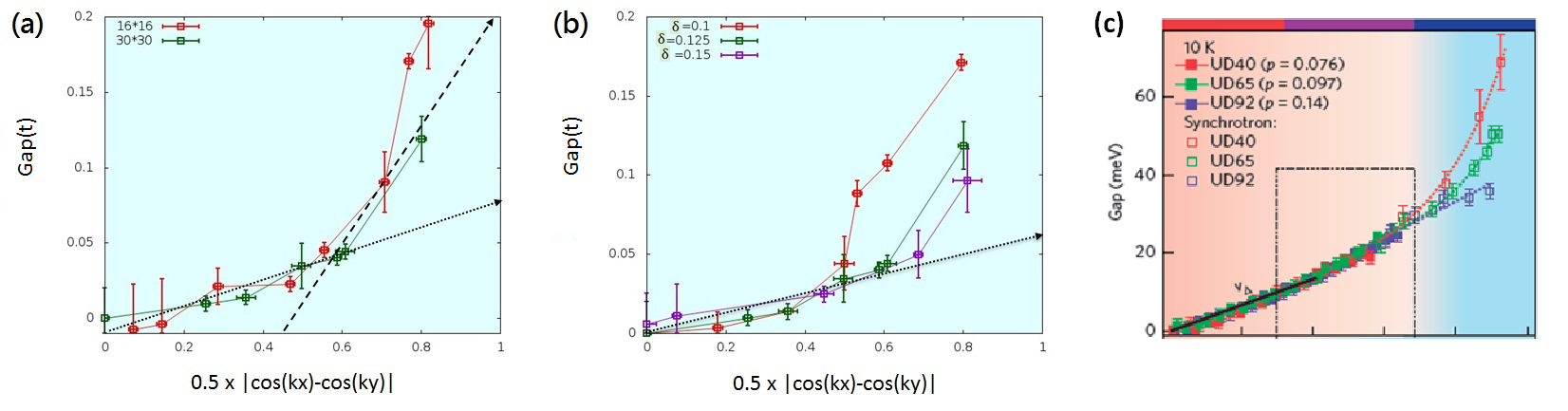}
\caption{}
\label{Fig.3}
\end{figure}

\begin{figure}[ht]
\centering
\includegraphics[width=\linewidth]{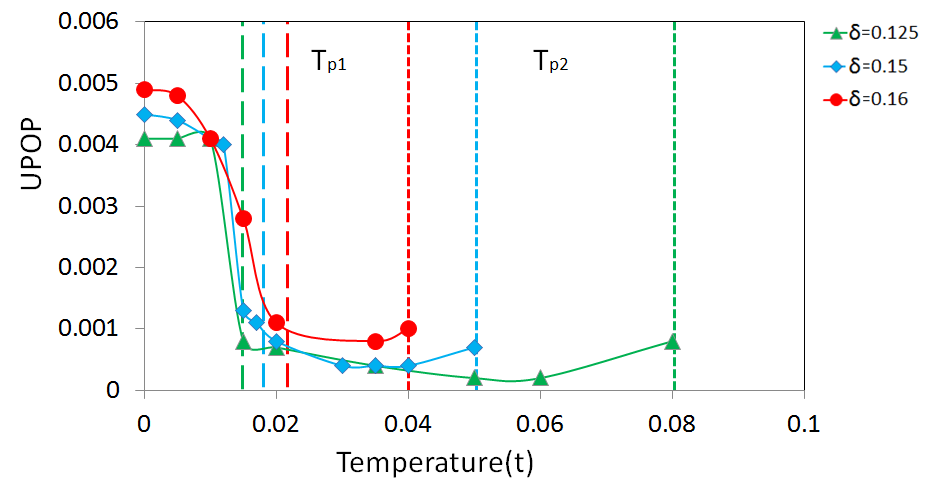}
\caption{}
\label{Fig.4}
\end{figure}

\begin{figure}[ht]
\centering
\includegraphics[width=\linewidth]{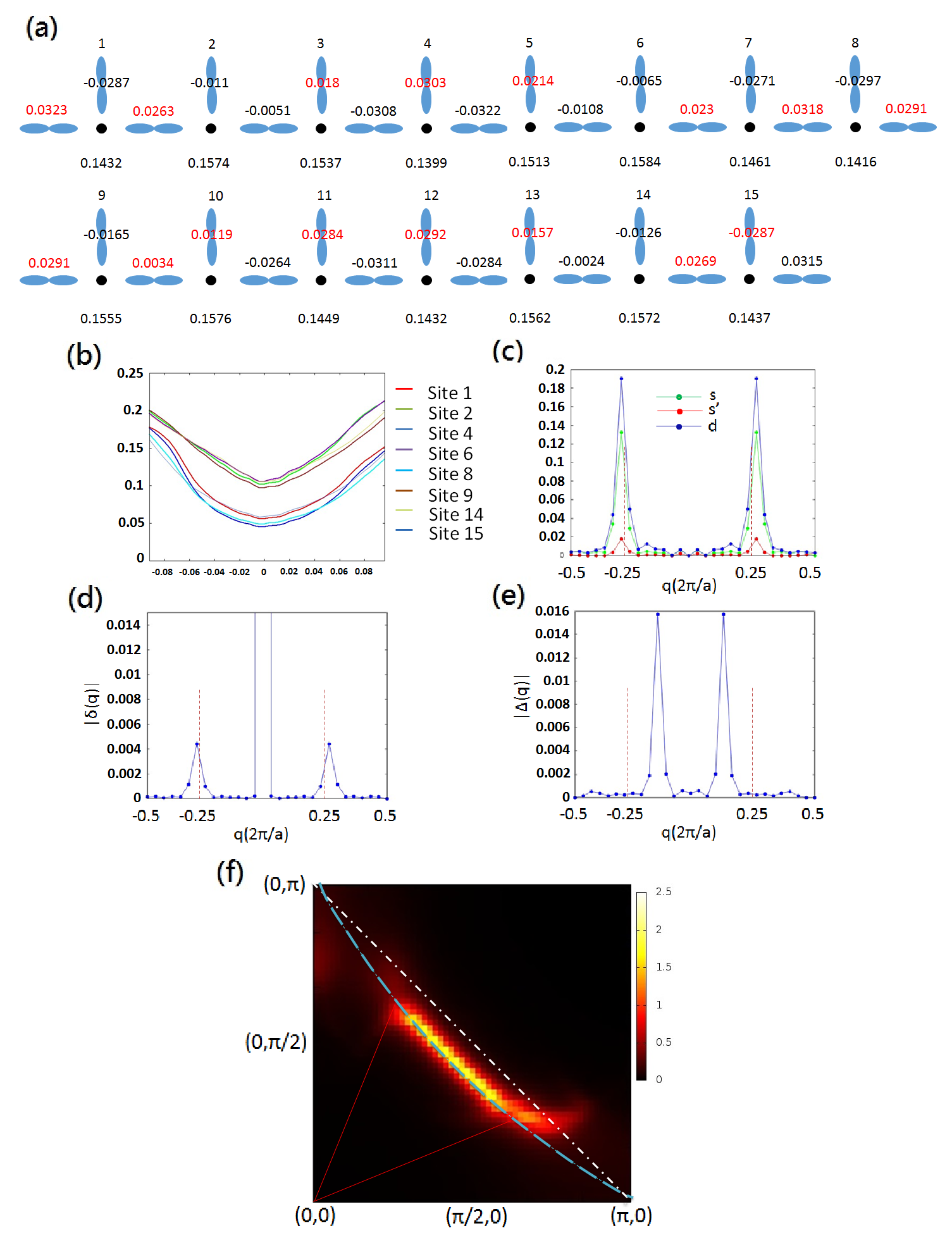}
\caption{}
\label{Fig.5}
\end{figure}

\begin{figure}[ht]
\centering
\includegraphics[width=\linewidth]{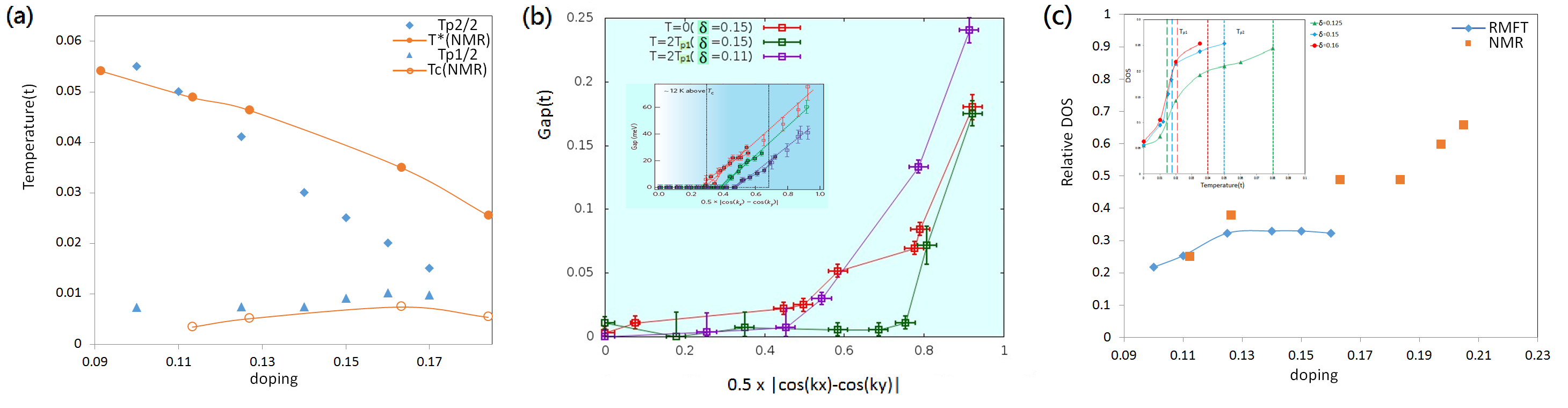}
\caption{}
\label{Fig.6}
\end{figure}

(a) Doping dependence of $T_{p1}$ and $T_{p2}$. $T_{p1}/2$ and $T_{p2}/2$ are shown with the blue triangles and diamonds respectively. The results from NMR \cite{Kawasaki} are also shown for comparison. We choose $0.1t \sim 464 K$. (b) The gap values scanned along the Fermi surface toward $(\pi,0)$ at $T=0$(red) and $2T_{p1}$(green) for $\delta=0.15$ and $2T_{p1}$(purple) for $\delta=0.11$. The inset shows the gap behavior of pseudo-gap phase from ARPES \cite{Hashimoto}. The meanings of symbols and colors are the same as Fig. \ref{Fig.3}(c). (c) Doping dependence of the relative DOS between IPDW and FLS(DOS$_{IPDW}$/DOS$_{FLS}$) at $T=0.035t$. The experimental data from \cite{Kawasaki} for $T=0$ is also plotted for comparison. The inset shows DOS of IPDW vs temperature for $\delta=0.125$, $0.15$, and $0.16$. $\Gamma$ we used here is $0.25 \sqrt{E^2+T^2}$ \cite{Alldredge}.

\clearpage



\section*{Supplementary material}

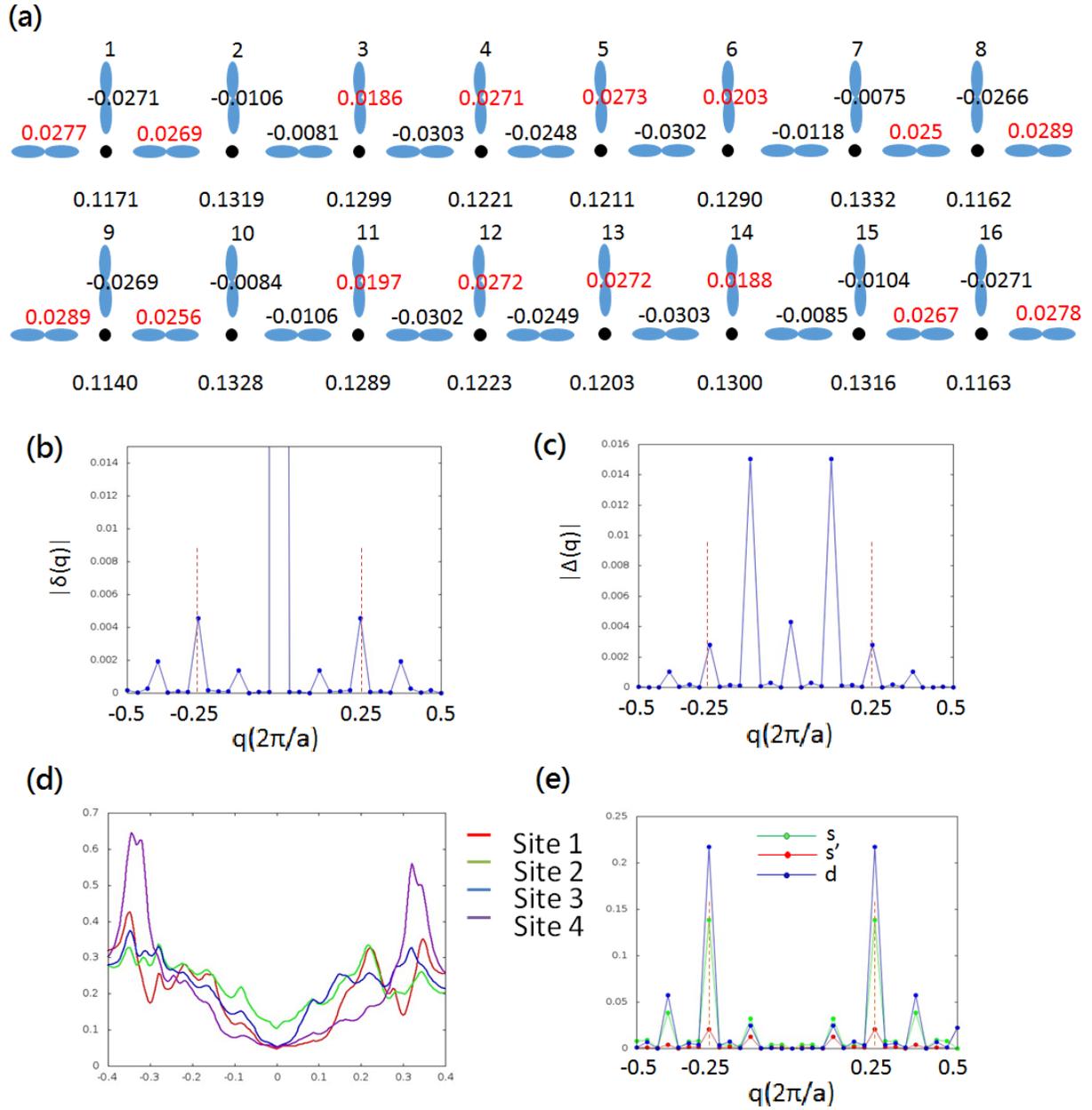

**Fig. S 1.** Properties of nPDW. (a) The real space modulation of nPDW in $32 \times 32$ lattice sites with $\delta = 0.125$. Since the pattern repeats itself with an inversion symmetry in the middle bond, here we only show the first 16 sites. The red and black numbers on each bond denote the values of pairing order and the number at each site (black dots) is the hole density. (b)(c) The Fourier transform of the value of hole density(b) and pairing order(c). (d) LDOS of the first 4 sites of this $32 \times 32$ nPDW. (e) Different form factors.

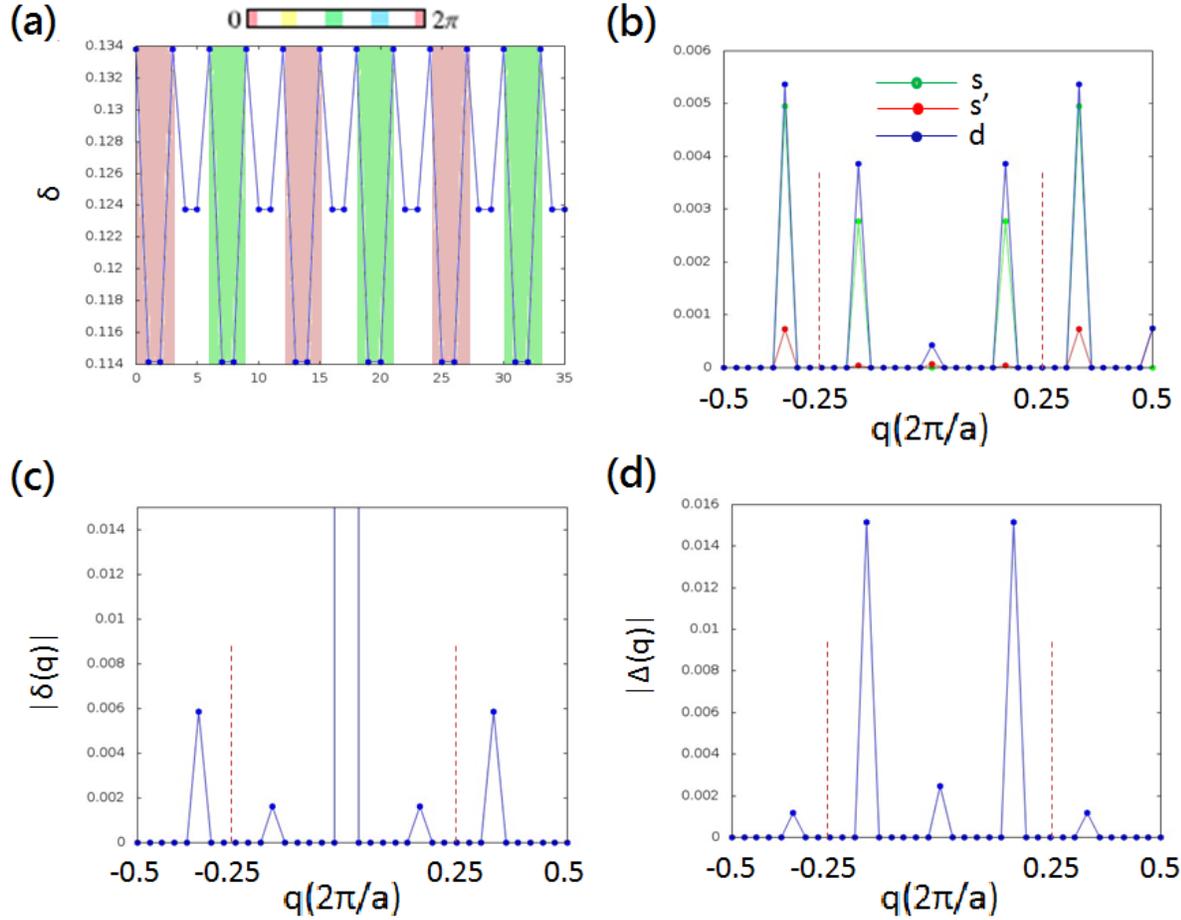

**Fig. S 2.** Figures showing the properties of discommensurate nPDW. (a) The phase variation of this pattern. Site 0-3, 12-15, and 24-27 are of phase equal to $0(2\pi)$ while sites 6-9, 18-21, and 30-33 are of phase $\pi$. (b) Form factors for discommensurate nPDW. We also include the Fourier transform of hole density(c) and pairing order(d).

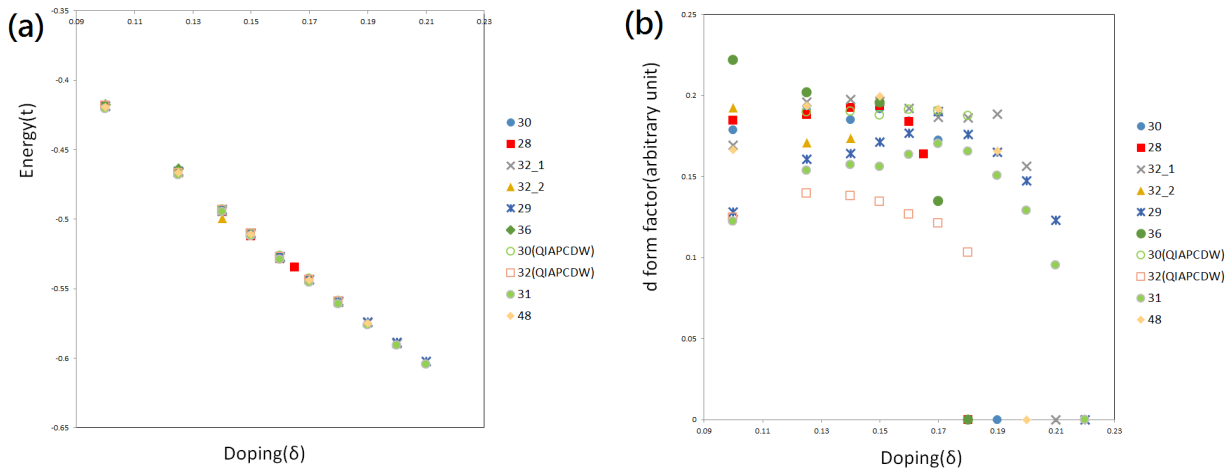

**Fig. S 3.** (a) Energies of several states chosen by us. Although we have listed ten different states here, their energies seem to be nearly degenerate and follow the same trend line. (b) Magnitude of $d$ form factor of patterns. Given different states we expect their magnitude to change but still all of them seem to have the same trend: the magnitude maintains the same until doping level exceeds 0.18, where it starts to drop drastically and becomes zero in the range of 0.18∼0.22.

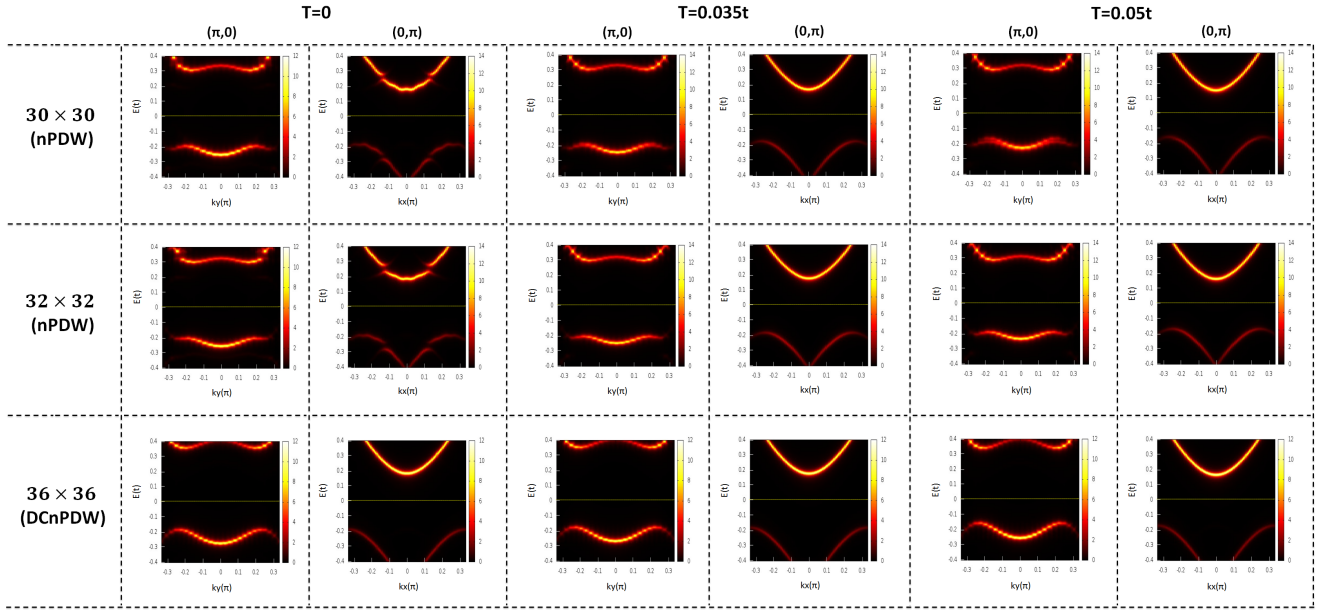

**Fig. S 4.** We list several quasi-particle spectra at antinodes($(\pi,0)/(0,\pi)$) for three different patterns at different temperatures. Although marked as nPDW in the first column, the patterns become IPDW at $T = 0.035t$ and $T = 0.05t$. However their spectra do not change much and the differences of gap values at $(\pi,0)$ and $(0,\pi)$ are within 10% [6].

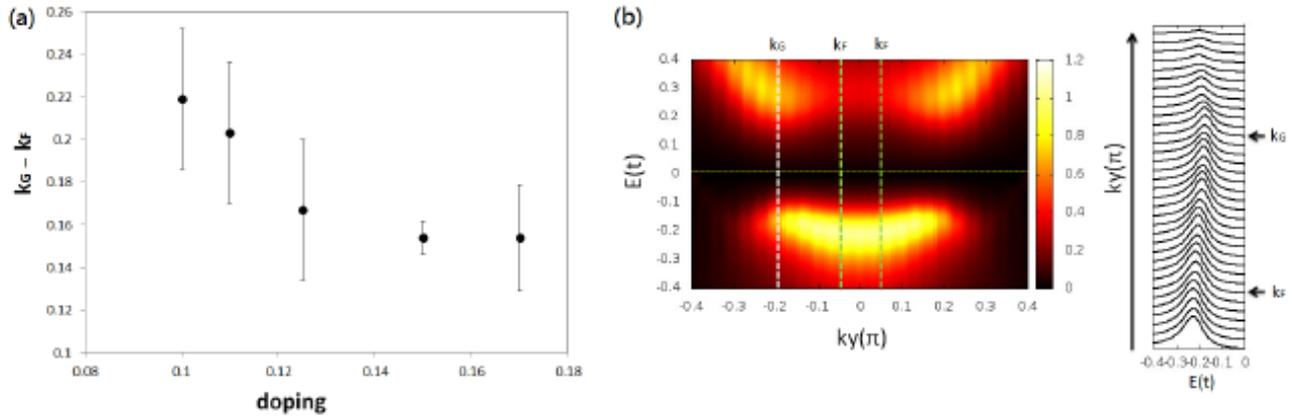

**Fig. S 5.** (a) A collection of several data points of $k_G - k_F$ vs doping at $k_x = \pi$. The way of determining the difference of $k_G$ and $k_F$ is shown in (b): $k_G$ determined by examining EDCs plotted from $k_y = 0$ toward $k_y = pi$, for dopant concentration 0.15. $k_F$ is determined by Fermi liquid surface and marked along with $k_G$ on the EDC plot. The quasiparticle spectra is also shown with Gaussian width $\Gamma = \alpha|E|$ ($\alpha = 0.25$) and marked with positions of $k_G$ and $k_F$.

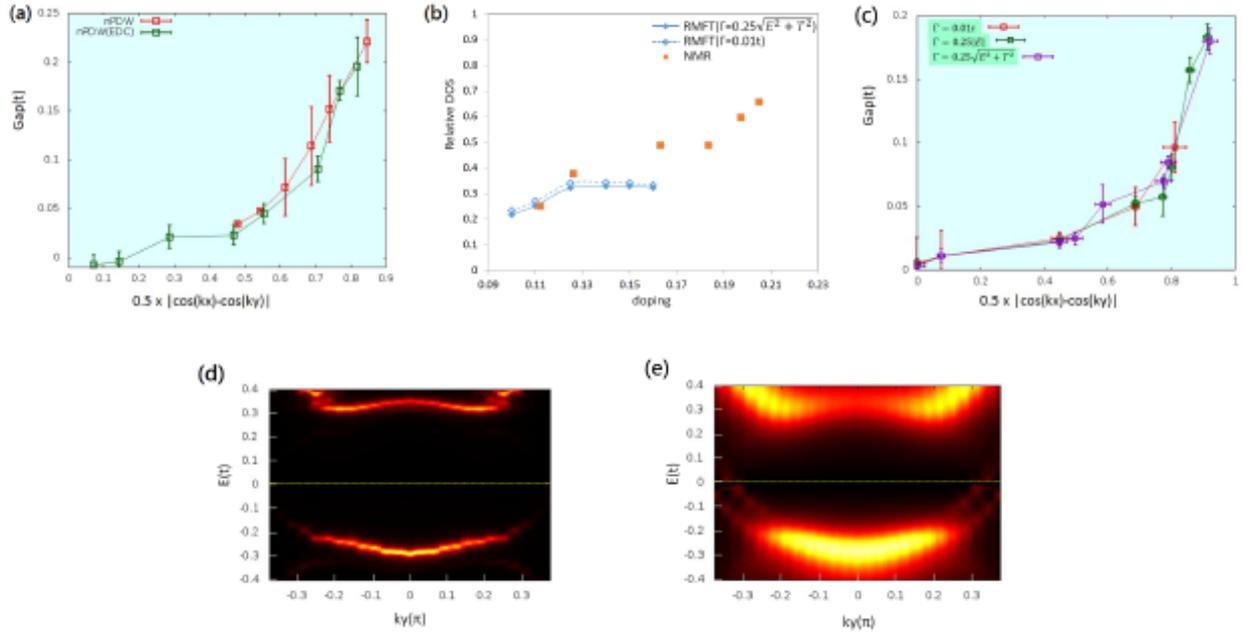

**Fig. S 6.** (a) Two-gap plot for nPDW at $\delta = 0.125$ as shown in Fig. 3 in the main text but obtained from different approaches: red line is determined by the gap values shown by quasiparticle spectra but green line comes from EDCs, the same way as in main text. (b) Relative DOS as a function of hole concentration as in Fig.7(b) in the main text but put together with two different $\Gamma$. The two blue lines are very close to each other. (c) Two gap plots determined by different $\Gamma$ for nPDW at $\delta = 0.15$. One can see that these lines nearly overlap with each other. Figure (d) and (e) again show the quasi-particle spectra for nPDW at $\delta = 0.125$ (for the $32 \times 32$ lattice) at $k_x = 0.977\pi$ but with different $\Gamma$: (d) $\Gamma = 0.01t$ and (e) $\Gamma = 0.25|E|$. Note that in fact (d) is identical as Fig. 1(f) in the main text. We can find that although these two figures look quite different due to the choices of $\Gamma$, important features such as location of $k_G$ are still the same, only that in (e) the spectra bands are broadened due to larger $\Gamma$.

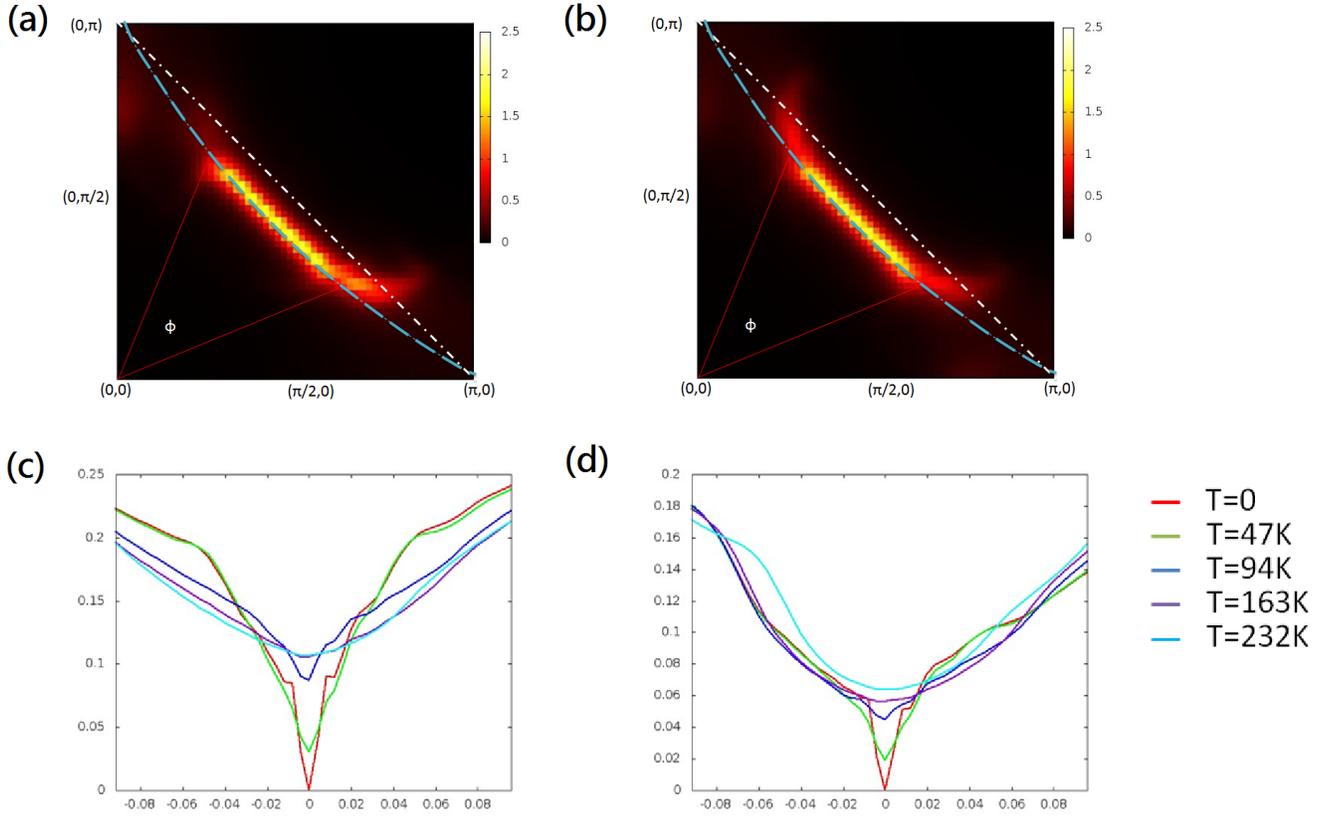

**Fig. S 7.** (a) and (b) Zero energy quasiparticle spectra in k space before(a) and after(b) taking average of x- and y-directions PDW. (a) is the same as Fig. 4f in the main text and we put it here again for the reason of comparison. Clearly, (b) looks more like the observation by experimental groups. (c) and (d) LDOS at sites near(c) and away from(d) domain walls at different temperatures for nPDW(IPDW) at $\delta = 0.15$. $\Gamma$ used here is equal to $\alpha\sqrt{E^2 + T^2}(\alpha = 0.25)$. All figures shown here are of $30 \times 30$ lattice size. Its $T_{p1}$ is around $90K$.

 

\end{document}